\documentclass{svproc}
\usepackage{url}

\usepackage{graphicx}

\begin{document}
\mainmatter              
\title{Uncertainties in the production of $p$ nuclides
in SN Ia
determined by Monte Carlo variations}
\titlerunning{Uncertainties in $p$ nuclide production
in SN Ia supernovae}  
%
\author{Thomas Rauscher\inst{1,2,9} \and Nobuya Nishimura\inst{3} \and Raphael Hirschi\inst{4,5,9} \and Alex St.J. Murphy\inst{6,9} \and Gabriele Cescutti\inst{7}
\and Claudia Travaglio\inst{8}}
\authorrunning{Thomas Rauscher et al.} 
%
%
\institute{Dept.\ of Physics, University of Basel, Switzerland,\\
\email{thomas.rauscher@unibas.ch}
\and
Centre for Astrophysics Research, University of Hertfordshire, UK
\and
Yukawa Institute for Theoretical Physics, Kyoto University, Japan
\and
Astrophysics Group, Faculty of Natural Sciences, Keele University, UK
\and
Kavli IPMU (WPI), University of Tokyo, Japan
\and
School of Physics and Astronomy, University of Edinburgh, UK
\and
INAF, Osservatorio Astronomico di Trieste, Italy
\and
INFN, Sezione di Torino, Italy
\and
UK Network for Bridging Disciplines of Galactic Chemical Evolution (BRIDGCE), \url{https://www.bridgce.net}
}

\maketitle              

\begin{abstract}
Several thousand tracers from a 2D model of a thermonuclear supernova were used in a Monte Carlo
post-processing approach to determine $p$-nuclide
abundance uncertainties originating from nuclear physics uncertainties in the reaction rates.
\end{abstract}
\section{Introduction}

Type Ia supernovae (SN Ia) originating from the explosion of a white dwarf accreting mass from a companion star have been suggested as
a site for the production of $p$ nuclides \cite{trav11}. The recently developed Monte Carlo (MC) code PizBuin \cite{rau16} was applied to the post-processing of temperature and density profiles obtained with tracer particles extracted from a 2D model of a thermonuclear supernova explosion. This code already has been applied to several other nucleosynthesis environments \cite{nob17,gab18} to tackle the question of how uncertainties in the nuclear reaction rates propagate into the final abundance yields. Realistic, temperature-dependent reaction rate uncertainties are used, combining experimental and theoretical uncertainties. Bespoke uncertainties are assigned to each individual rate and all rates are varied simultaneously within their uncertainty limits. This approach allows to probe the combined action of all uncertainties and proved superior to manual variation of a few rates or coupled variation of rate subsets.

In this study, 51200 tracers were extracted from the DDT-a explosion model as described in \cite{trav11}. Among these, 4624 tracers experienced conditions supporting the production of $p$ nuclides and their temperature and density profiles were used in the MC
post-processing. The reaction network included 1342 nuclides (around stability and towards the proton-rich side). To complete the study it had to be run more than 40 million times. This necessitated the use of HPC facilities.

\section{Results}

\begin{figure}
\includegraphics[width=\columnwidth]{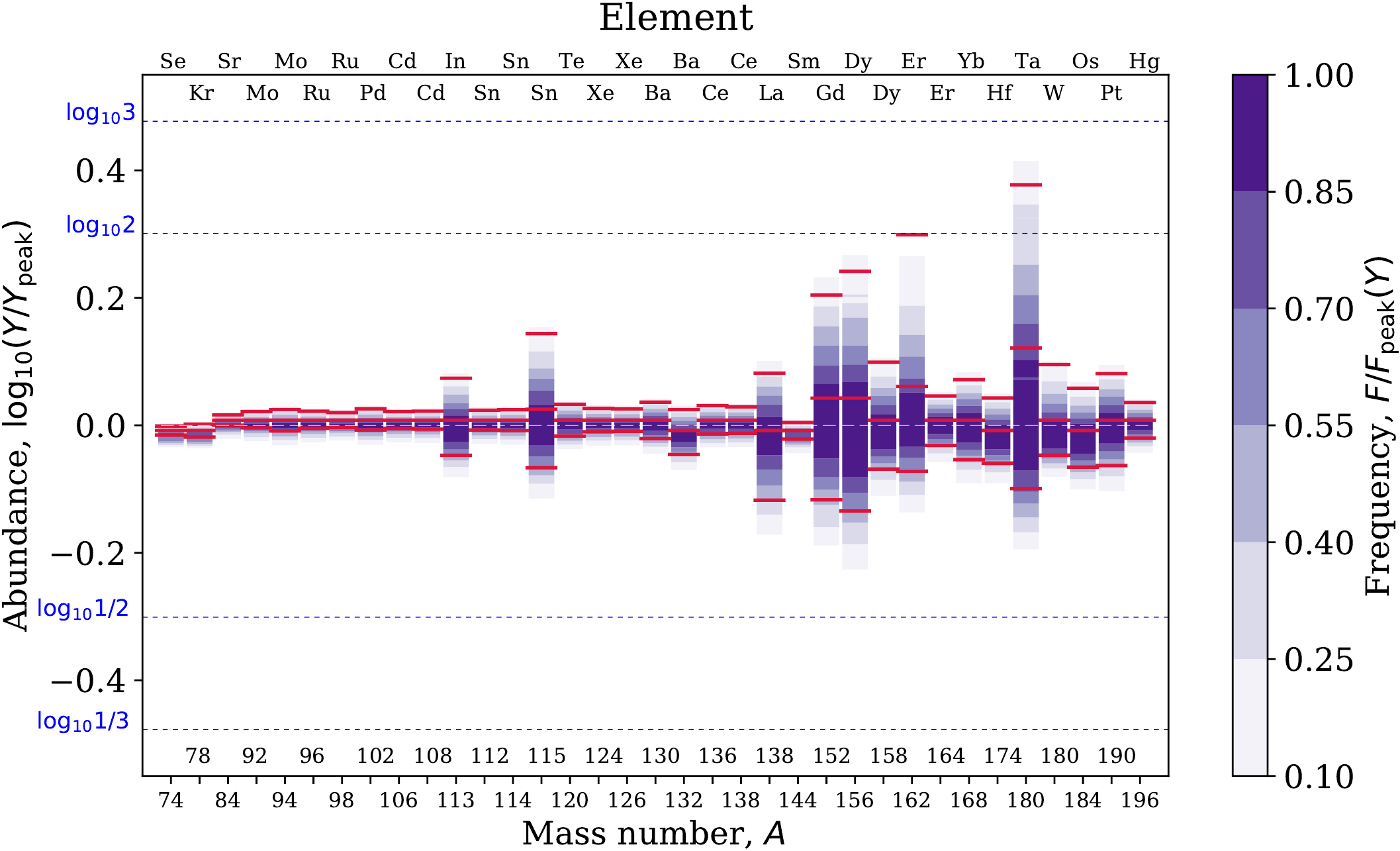}
\caption{Total production uncertainties of $p$ nuclide due to rate uncertainties. The color shade gives the relative probabilistic frequency and the horizontal red lines enclose a 90\% probability interval for each nuclide. Uncertainty factors of two and three are marked by dotted lines. Note that the uncertainties are asymmetric and that the abundance scale is logarithmic. [Figure taken from \cite{nob18}, with permission.] \label{fig:uncertall}}
\end{figure}

Figure \ref{fig:uncertall} shows the total production uncertainties (all tracers combined) for each $p$ nuclide. With the exception of $^{180}$Ta, which is known to receive major contributions from other nucleosynthesis processes, the uncertainties are well below a factor of two, despite the fact that photodisintegration, electron capture, and $\beta^+$-decay rates of unstable nuclides bear much larger uncertainties. The uncertainties are also considerably smaller than those found for the production of $p$ nuclides in the $\gamma$-process in explosions of massive stars (core-collapse supernovae, ccSN) \cite{rau16}. This can be explained by the larger number of temperature-density combinations encountered in SN Ia, which allow alternative reaction flows bypassing suppressed reactions.

Due to the challenging demand on CPU time, only one SN Ia explosion model was studied. To be able to draw more general conclusions, uncertainty contributions from high and low density
regions in the white dwarf were also scrutinised separately. The high density regions gave rise to larger uncertainties in the final abundances \cite{nob18}. Based on the ratio of high- to low-density regions in other models, our results can be used to estimate the resulting uncertainties also in those other models.

As in our previous investigations \cite{rau16,nob17,gab18}, key rates were identified by correlations between rate and abundance variations. Only one reaction was found to dominate the total production uncertainty: The uncertainty in $^{145}$Eu$+\mathrm{p} \leftrightarrow \gamma + ^{146}$Gd significantly affects the abundance uncertainty of $^{146}$Sm. Again, this is due to the range of conditions found in SN Ia. Considering high- and low-density regions separately, a few other key reactions were identified. For the low-density group, five key rates were found:  $^{129}$Ba$+\mathrm{n} \leftrightarrow \gamma + ^{130}$Ba for $^{130}$Ba, $^{137}$Ce$+\mathrm{n} \leftrightarrow \gamma + ^{138}$Ce for $^{138}$Ce, $^{144}$Sm$+\alpha \leftrightarrow \gamma + ^{148}$Gd for $^{146}$Sm, $^{164}$Yb$+\alpha \leftrightarrow \gamma + ^{168}$Hf for $^{168}$Yb, and $^{186}$Pt$+\alpha \leftrightarrow \gamma + ^{190}$Hg for $^{190}$Pt.
For the high-density group, seven key rates were identified: $^{83}$Rb$+\mathrm{p} \leftrightarrow \gamma + ^{84}$Sr for $^{84}$Sr, $^{105}$Cd$+\mathrm{n} \leftrightarrow \gamma + ^{106}$Cd for $^{106}$Cd, $^{111}$Sn$+\mathrm{n} \leftrightarrow \gamma + ^{112}$Sn for $^{112}$Sn, $^{129}$Ba$+\mathrm{n} \leftrightarrow \gamma + ^{130}$Ba for $^{130}$Ba, $^{137}$Ce$+\mathrm{n} \leftrightarrow \gamma + ^{138}$Ce for $^{138}$Ce, $^{176}$W$+\alpha \leftrightarrow \gamma + ^{180}$Os for $^{180}$W, and $^{186}$Pt$+\alpha \leftrightarrow \gamma + ^{190}$Hg for $^{190}$Pt.

For further details on calculation and results, see \cite{nob18}.

\paragraph{Acknowledgments}
This work has been partially supported by the European Research Council
(EU-FP7-ERC-2012-St Grant 306901), the EU COST Action CA16117 (ChETEC),
the UK STFC (ST/M000958/1), and MEXT Japan (Priority Issue on Post-K computer: Elucidation of the Fundamental Laws and Evolution of the Universe).
Parts of the computations were carried out on
COSMOS (STFC DiRAC Facility) at the University of Cambridge. Further computations were carried out at CfCA, NAO Japan, and at YITP, Kyoto University.

%
%


\begin{thebibliography}{99}
%
%
\bibitem{trav11}
Travaglio, C., R\"opke, F. K., Gallino, R., Hillebrandt, W.: Type Ia Supernovae as Sites of the p-process: Two-dimensional Models Coupled to Nucleosynthesis.
Ap. J. 739, 93 (2011). \url{doi:10.1088/0004-637X/739/2/93}

\bibitem{rau16}
Rauscher, T., Nishimura, N., Hirschi, R., Cescutti, G., Murphy, A. St.J., Heger, A.: Uncertainties in the production of $p$ nuclei in massive stars obtained from Monte Carlo variations.
MNRAS 463, 4153 (2016). \url{doi:10.1093/mnras/stw2266}

\bibitem{nob17}
Nishimura, N., Hirschi, R., Rauscher, T., Murphy, A. St.J., Cescutti, G.: Uncertainties in $s$-process nucleosynthesis in massive stars determined by Monte Carlo variations.
MNRAS 469, 1752 (2017). \url{doi:10.1093/mnras/stx696}

\bibitem{gab18}
Cescutti, G., Hirschi, R., Nishimura, N., den Hartogh, J. W., Rauscher, T., Murphy, A. St.J, Cristallo, S.: Uncertainties in s-process nucleosynthesis in low-mass stars determined from Monte Carlo variations.
MNRAS 478, 4101 (2018). \url{doi:10.1093/mnras/sty1185}

\bibitem{nob18}
Nishimura, N., Rauscher, T., Hirschi, R., Murphy, A. St.J., Cescutti, G., Travaglio, C.: Uncertainties in the production of p nuclides in thermonuclear supernovae determined by Monte Carlo variations.
MNRAS 474, 3133 (2018). \url{doi:10.1093/mnras/stx3033}



\end{thebibliography}
\end{document}